\documentclass[showpacs,prb,12pt,onecolumn]{revtex4-1}

\usepackage{graphicx}%
\usepackage{amssymb}

\begin{document}\normalsize       

\date{\today}

\title{\bf The examination of stable charge states of vacancies in Cu$_2$ZnSnS$_4$}
\author{Xiaoli Zhang}
\affiliation{Key Laboratory of Materials Physics, Institute of Solid
State Physics, Chinese Academy of Sciences, Hefei 230031, China}
\author{Miaomiao Han}
\affiliation{Key Laboratory of Materials Physics, Institute of Solid
State Physics, Chinese Academy of Sciences, Hefei 230031, China}
\author{Zhi Zeng}
\email{zzeng@theory.issp.ac.cn} \affiliation{Key Laboratory of
Materials Physics, Institute of Solid State Physics, Chinese Academy
of Sciences, Hefei 230031, China} \affiliation{University of Science
and Technology of China, Hefei 230026, China }
\author{Xiaoguang Li}
\affiliation{Key Laboratory of Materials Physics, Institute of Solid
State Physics, Chinese Academy of Sciences, Hefei 230031, China}
\affiliation{University of Science and Technology of China, Hefei
230026, China }
\author{H. Q. Lin}
\affiliation{Beijing Computational Science Research Center, Beijing
100084, China}

\date{\today}

\begin{abstract}
The stable charge states of vacancies in the solar cell absorber
material Cu$_2$ZnSnS$_4$ are investigated using Kohn-Sham (KS)
defect-induced single particle levels analysis by concerning the
screened Coulomb hybrid functional. We find out that the Cu, Zn and
S vacancies (denoted by V$_{Cu}$, V$_{Zn}$, V$_{S}$) do not induce
single particle defect levels in the vicinity of the band gap thus
each of them has only one stable charge state corresponding to the
fully occupied valence band V$_{Cu}^{1-}$, V$_{Zn}^{2-}$ and
V$_{S}^{0}$, respectively (and therefore cannot account for any
defect transition energy levels). The Sn vacancy (V$_{Sn}$) has
three stable charge states V$_{Sn}^{2-}$, V$_{Sn}^{3-}$ and
V$_{Sn}^{4-}$, which may account for two charge transition energy
levels. By comparing with previous charge transition energy levels
studies, our results indicate that the examination of stable charge
states is a necessary and important step which should be done before
charge transition energy levels calculations.
\end{abstract}

\maketitle
\newpage

\section{\bf INTRODUCTION}
Cu(In,Ga)Se$_2$ (CIGS) possesses the highest energy conversion
efficiency up to 20.3$\%$ among the thin film solar
cells\cite{Jackson}. Researchers are still making their efforts on
improving the CIGS solar cell efficiency\cite{Chirila,Contreras}.
CIGS solar cells compete today as successors of the dominating
silicon technology. Nevertheless, there are concerns about their
large scale production due to the price and the availability of In.
By replacing In and Ga elements in CIGS with Zn and Sn elements it
formulates Cu$_2$ZnSnSe$_4$ (CZTSe)\cite{Chen1}, which is very
similar to Cu$_2$ZnSnS$_4$ (CZTS) in structure and physical
properties\cite{Chen2,Perssona}. CZTS and CZTSe quaternary compounds
overcome the disadvantages in CIGS owing to their earth-abundant and
low-toxicity constituents. Moreover, the kesterite CZTS  possesses
the \textit{I}$\bar{4}$ space group\cite{Hall}, characteristic
optimal band gap of about 1.5 eV$\cite{Seol,Scragg}$ (bandgap of
CZTSe is 1.0 eV) and large absorption coefficient of more than
10$^4$ cm$^{-1}.$\cite{Ito} The energy conversion efficiency of
CZTS-based solar cell has achieved a world record of 12.6$\%$ in
November 2013\cite{Wang}. The relatively high energy conversion
efficiency is partly due to the \textit{p}-type conductivity via
inducing the intrinsic point defects in the parent CZTS. Point
defects significantly influence the efficiency of the
\textit{pn}-junction-based solar cells. Therefore, much attention
has been paid to the point defects in CZTS to understand the
relationship between defects and solar cell
efficiency.$\cite{Zillner,Washio,Chen3,Walsh,Nagoya,Han}$.

Previous theoretical researches mainly focus on the formation
energies and defect transition energy levels of the intrinsic
defects in CZTS using the first-principles total energy
calculations$\cite{Chen3,Walsh,Nagoya,Han}$ based on density
functional theory\cite{Hohenberg,Kohn}. Formation energies define
the formation ability of defects in samples. The defects with low
formation energies can exist in large amount in the samples while
those with high formation energies can exist in relatively small
amount which will not affect the properties of the CZTS. Defect
transition energy levels are directly related to the defect (donor
or acceptor) levels in the band gap. Deep defect levels mainly act
as recombination centers, while shallow defect levels allow the
transformation from one charge state to another easily which will
increase carrier concentration in the sample. The theoretical value
of a defect transition energy level is closely related to the total
energy difference of two charge states of the defect, which was
always taken in previous studies.$\cite{Chen3,Walsh,Nagoya,Han}$
However, the way of inspecting total energy difference is not able
to examine the stability of the considered charge states, which may
lead to theoretically unphysical defect transition energy
levels$\cite{Castleton,Oikkonen1,Oikkonen2,Pohl}$. Taking Cu vacancy
(V$_{Cu}$) as an example, first, the neutral V$_{Cu}^{0}$ was
simulated by removing an Cu atom in the supercell, and the
V$_{Cu}^{1-}$ charge states was simulated by adding an additional
electron to the jellium background of V$_{Cu}^{0}$. Next, the two
models were fully relaxed to get the total energy difference and the
value of defect transition energy level $\epsilon$(-/0). In this
procedure one ignored the examination of whether the 0 and 1- charge
states are stable, which should be done before defect transition
energy level calculation$\cite{Castleton,Oikkonen1,Oikkonen2,Pohl}$.
Therefore, in this paper, we are motivated to go a prior step to
identify the possible stable charge states of defects in CZTS using
band structures and defect-induced single particle levels (Kohn-Sham
eigenvalues) analysis$\cite{Castleton,Oikkonen1,Oikkonen2}$.

To identify the possible stable charge states of all the intrinsic
defect completely, one would better investigate all vacancies,
antisites, intersites and defect complexes in the CZTS. Due to the
bigger defect number in CZTS than in CuInSe$_2$, in this paper, only
the vacancies are considered as a first step to explore the possible
stable charge states of defects in CZTS.

Our results show that all the enforced charge states of Cu vacancy
V$_{Cu}$ and Zn vacancy V$_{Zn}$ and S vacancy V$_{S}$ cannot
produce localized defect-induced single particle levels in the
vicinity of band gap as well as defect transition energy levels
within the band gap, for that each of them only has one stable
charge state corresponding to the fully occupied valence band
V$_{Cu}^{1-}$, V$_{Zn}^{2-}$ and V$_{S}^{0}$, respectively. The
exception is that V$_{Sn}$ (Sn vacancy) in its 2-, 3-, 4- charge
states are stable due to that the  defect-induced single particle
levels are located at the top of the valence band maximum (VBM).
Consequently, Cu, Zn and S vacancies can not while Sn vacancy can
induce charge transition energy levels in the band gap, which is
partly at odd with previous charge transition energy levels studies.
Our results indicate that the examination of stable charge states is
a necessary and important step which should be done before charge
transition energy levels calculations.

\section{\bf Theoretical approaches}
The electronic structure calculations are carried out using the
density functional theory as implemented in the plane wave VASP
code$\cite{Kresse}$. For the exchange-correlation functional, both
generalized gradient approximation (GGA) of Perdew-Burke-Ernzerhof
(PBE)\cite{Kumagai,Perdew} and screened Coulomb hybrid functional
Heyd-Scuseria-Ernzerhof (HSE06) are used $\cite{Heyd,Heyd2}$.
PBE-GGA is used to obtain an initial structures and HSE06 is used to
calculate the electronic structures, since GGA often under estimates
the band gap of Cu-based semiconductor compounds and HSE06
functional describes the localized orbitals more correctly than
(semi)local-density functionals$\cite{Henderson}$ by substituting
part of short range PBE-GGA exchange energy with the short range
Hartree-Fock (HF) exchange energy. HSE06 functional is verified to
improve the description of the band gap of CZTS as well as the
defect properties$\cite{Han,Chen2}$. The parameter controlling the
amount of HF exchange in the HSE06 functional is set to 0.3 and the
range-separation screening parameter $\mu$ is set to 0.20
{\AA}$^{-1}$. The interaction between ions and electrons is
described by the projector-augmented wave (PAW)
method$\cite{Bloch,Kresse2}$. The cutoff energy for the plane-wave
basis is set to 300 eV. Using these setup our calculated band gap of
1.43 eV is close to the reported 1.44 - 1.51 eV for CZTS
$\cite{Seol,Scragg}$. The vacancy models are constructed by removing
an atom from a 64-atom 2$\times$2$\times$1 supercell.  The Brillouin
zone sampling is done using 2$\times$2$\times$2 \textit{k}-point
mesh. The ground state geometries of all the defect systems are
obtained by minimizing the Hellman-Feynman forces on each atom to
become less than 0.02 eV/{\AA} .

A defect often has multiple possible charge states. Taking V$_{Sn}$
as an example, V$_{Sn}$ has five possible charge states from -4 to
0. The stability of these possible charge states is strongly related
to the positions of the defect-induced single particle levels
relative to the band gap. On one hand, if a defect with a certain
charge state induces single particle levels in the vicinity of the
band gap, the corresponding charge state is stable because the
defect-induced single particle levels may form a stable bound state
bounding electrons or holes. On the other hand, if a defect with a
certain charge state induces single particle levels in the valence
band or conduction band, the corresponding charge state is unstable
because the defect-induced single particle levels are always
occupied or unoccupied and the electrons or holes on them are
communized. A defect with multiple stable charge states is a
necessary though not a sufficient condition for transitions among
the stable charge states to take place. Such transitions produce
defect transition energy levels as others investigations by DFT
calculations. Additionally, a defect with all the charge states
produce none single particle levels in the vicinity of the band gap
will stable at one charge state corresponding to the fully occupied
valence band. Other charge states are physically unstable because
the existence of either holes at the VBM or electrons at the
conduction band maximum (CBM)\cite{Castleton}. This kind of defects
will not have charge transition energy level in the gap. Therefore,
we conclude that the examination of possible stable charge states
must be done before charge transition energy levels calculations. To
pick out the defect-induced single levels, we examine the local
density of states (LDOS) of vacancy systems which corresponds to the
DOS of the four nearest neighbor atoms around the vacant site. The
additional sharp peaks in density of states (LDOS) of a vacancy
system compared with that of parent compound are defect-induced
single particle levels.

It should be emphasized that this work provides a qualitative view
of the basic nature of defect in CZTS, which is also a precondition
of the quantitative view from defect transition energy levels
calculations. The qualitative investigation of CuInSe$_2$ was
previously done by Oikkonen et. al, which lead to new insight about
the defect physics in CuInSe$_2$. Unlike total energy difference
investigations which emphasize the importance of V$_{Cu}$, they
found out that only Se related defects can induce charge transition
energy levels in the gap. Hopefully, our work may improve the
understanding of the defect physics in CZTS.

\section{\bf RESULTS AND DISCUSSION}
The movements of atoms of CZTS induced by the vacancy under the
relaxation are qualitatively similar for Cu, Zn and Sn vacancy
(denoted as V$_{Cu}$, V$_{Zn}$ and V$_{Sn}$) as shown in Fig. 1,
which can be understood by the similar surroundings of Cu, Zn and Sn
vacancies (four nearest-neighbor S atoms ). The four
nearest-neighbor S atoms move toward the Cu (Zn and Sn) vacancy, and
therefore decrease the V$_{Cu}$-S (V$_{Zn}$-S and V$_{Sn}^0$-S) bond
with 0.04 {\AA} (0.005 {\AA} and 0.14 {\AA}) compared with the
parent compound. The relaxation difference of the nearest S atoms
around the vacancies is determined by the size mismatch between Cu,
Zn and Sn ions. It is obvious that the
 displacements of four nearest-neighbor S atoms around
V$_{Sn}$ are decreasing with increasing electronnegativity, which
can be explained by the increasing Coulomb repulsion among the four
nearest-neighbor S atoms. In the kesterite CZTS, the S atom has four
nearest neighbors: two Cu, one Zn and one Sn atoms. With the
creation of a S vacancy, the nearest neighbor Cu and Zn atoms move
away from the S vacancy site, whereas the Sn atoms relax toward the
vacancy site.

To figure out the defect-induced single particle levels of the
V$_{Cu}$ and V$_{Zn}$, we look into the LDOS of the four nearest
neighbor S atoms around the V$_{Cu}$ and V$_{Zn}$ vacancy,
respectively. Our KS band structure analysis shows that all the
charge states of V$_{Cu}$ and V$_{Zn}$ can not produce
defect-induced single particle levels in the vicinity of the band
gap. Consequently, each Cu or Zn vacancy has only one stable charge
state corresponding to the fully occupied valence bands
V$_{Cu}^{1-}$ or V$_{Zn}^{2-}$, seeing Fig. 2. where the
defect-induced single particle levels are marked in the band
structures with red lines. The defect-induced single particle levels
introduced by V$_{Cu}^{1-}$ and V$_{Zn}^{2-}$ locate deep at about
2.7 eV below the VBM, as shown in Fig. 2. For V$_{Zn}^{2-}$,
additional two defect-induced single particle levels lying right
below the VBM. Such defect-induced single particle levels lying in
the valence band are always occupied, and will not influence the
electric behavior of CZTS, which could be seen from the almost
unchanged band gap of V$_{Cu}^{1-}$ and 0.1 eV band gap decrease of
V$_{Zn}^{2-}$ in comparing to the parent material in Fig. 2.

Contrary to the V$_{Cu}$ and V$_{Zn}$ vacancies, the V$_{Sn}$ with
multiple charge states create defect-induced single particle levels
at the VBM as shown in Fig. 2. The neutral V$_{Sn}^{0}$ creates two
defect levels lying at about 0.5 eV below the VBM, but the addition
of 2 to 4 electrons forming 2- to 4- charge states shift one level
up to the top of the valence band, which can be explained by the
increasing levels Coulomb repulsion. The shifting of the defect
states is accompanied by the reduced lattice relaxation around the
vacant sites. As shown in Fig. 1, the four nearest neighbor S atoms
move towards the vacant sites for V$_{Sn}^{1-}$, V$_{Sn}^{2-}$ and
V$_{Sn}^{3-}$ with decreasing displacement, and move away from the
vacant site for V$_{Sn}^{4-}$. The shallow defect-induced single
particle levels of V$_{Sn}$ significantly change the electronic
structure of CZTS. It is obvious that the band gaps decrease of
0.23, 0.36, 0.48, 0.61 and 0.76 eV for charge states from 0 to 4-,
respectively, compared with parent CZTS. The defect levels at the
top of the VBM may behave as an acceptor level if holes can be
thermally excited to the VBM at room temperature, thereby
introducing a shallow acceptor level in the CZTS gap. However,
previous theoretical studies show that the V$_{Sn}$ has high
formation energy and high charge transition levels and will not be a
effective \textit{p}-type acceptor.

Unlike cation vacancies, all the enforced charge states for anion S
vacancy do not induce defect levels in the interested energy region.
This is in consistent with HSE06 findings by D. Han \textit{et
al.}$\cite{Han}$, who found out that there is no defect transition
level for V$_S$ in the band gap. If one charging two additional
electrons into the neutral charge state V$_S^0$, the excess charge
does not fully localize on the defect but fills the conduction band.
The band structure and LDOS of V$_S^{2+}$ charge state is also
examined, which show VBM up shift crossing the Fermi level.
Therefore, the neutral V$_S^0$ is the only stable charge state for
V$_S$, which corresponds to the fully occupied valence band. The
characteristic of anion S vacancy in CZTS is also very different
from that of Se vacancy in CuInSe$_2$. Se vacancies can produce
defect-induced single particle levels near the band gap region in
CuInSe$_2$$\cite{Oikkonen1,Oikkonen2}$. We guess it might be caused
by the difference of intrinsic characteristics between Se and S
atoms, for example, lower $\textit{p}$ orbital energy level and
larger atomic size of Se than that of S atom$\cite{Chen}$. To verify
our conjecture, we investigate the Se vacancies in the CZTSe.
However, no single particle levels appear near the band gap region.
Therefore, the difference of vacancies electron structures between
CZTS and CuInSe$_2$ is caused by the different cations. This in turn
implies that the Sn atoms significantly affect the electron
structure of CZTS.

In summary, our results conclude that Cu, Zn and S vacancies have
only one stable charge states and therefore can not induce charge
transition levels in the band gap. Moreover, Sn vacancy with 2-, 3-,
and 4- charge states are stable which indicates that two charge
transition energy levels corresponding to $\epsilon$(3-/4-) and
$\epsilon$(3-/2-) might be induced in the band gap. Our results are
very different from previous charge transition energy level
investigations which show that each Cu, Zn and S vacancy has one
charge transition energy level, while Sn vacancy has three charge
transition energy levels$\cite{Chen3}$. We think that the more
charge transition energy levels obtained from previous charge
transition energy level study compared with our results are caused
by the consideration of the unstable charge states. Therefore, our
conclusion in turn indicate that the examination of stable charge
states is a necessary and important step which should be done before
charge transition energy levels calculations.

\section{\bf CONCLUSIONS }
In this work, we investigated the band structures of the four
vacancy defects in CZTS by employing screened Coulomb hybrid
functional Heyd-Scuseria-Ernzerhof (HSE06). Our results reveal that
the Sn vacancy with 2-, 3-, and 4- charge states can provide
Kohn-Shan (KS) defect levels in the vicinity of band gap while Cu,
Zn and S vacancies can not. Thus Cu, Zn and S vacancies have only
one stable charge states corresponding to V$_{Cu}^{1-}$,
V$_{Zn}^{2-}$ and V$_{S}^{0}$, respectively, which can not induce
charge transition energy level in the band gap. Our results are
partly at odd with previous charge transition energy levels studies,
however, indicate that the examination of stable charge states is a
necessary and important step which should be done before charge
transition energy levels calculations.

\section{\bf ACKNOWLEDGMENTS}
This work was supported by the special Funds for Major State Basic
Research Project of China (973) under Grant No. 2012CB933702, the
NSFC under Grant Nos. 11204310 and U1230202(NSAF), Heifei Center for
Physical Science and Technology under Grant No. 2012FXZY004. The
calculations were performed in Center for Computational Science of
CASHIPS and on the ScGrid of Supercomputing Center, Computer Network
Information Center of CAS.


\clearpage

\begin{figure}[tp]
\vglue 1.0cm
\newpage


\caption {(color online). Schematic representation of atomic
relaxation (expressed in ${\AA}$) around each vacancy.}


\caption {(color online). The band structure of parent CZTS,
V$_{Cu}^{1-}$, V$_{Zn}^{2-}$,  V$_{S}^{0}$ and Sn vacancy with 0,
1-, 2-, 3-, 4- four different charge states calculated in a 64-atom
supercell. The dashed lines illustrate the defect levels induced by
the vacancy compared to the parent band structure in each case.}
\end{figure}


\begin{figure*}[htbp]
\center {$\Huge\textbf{Fig.1  \underline{Zhang}.eps}$}
\includegraphics[bb = 10 10 700 700, width=1.2\textwidth]{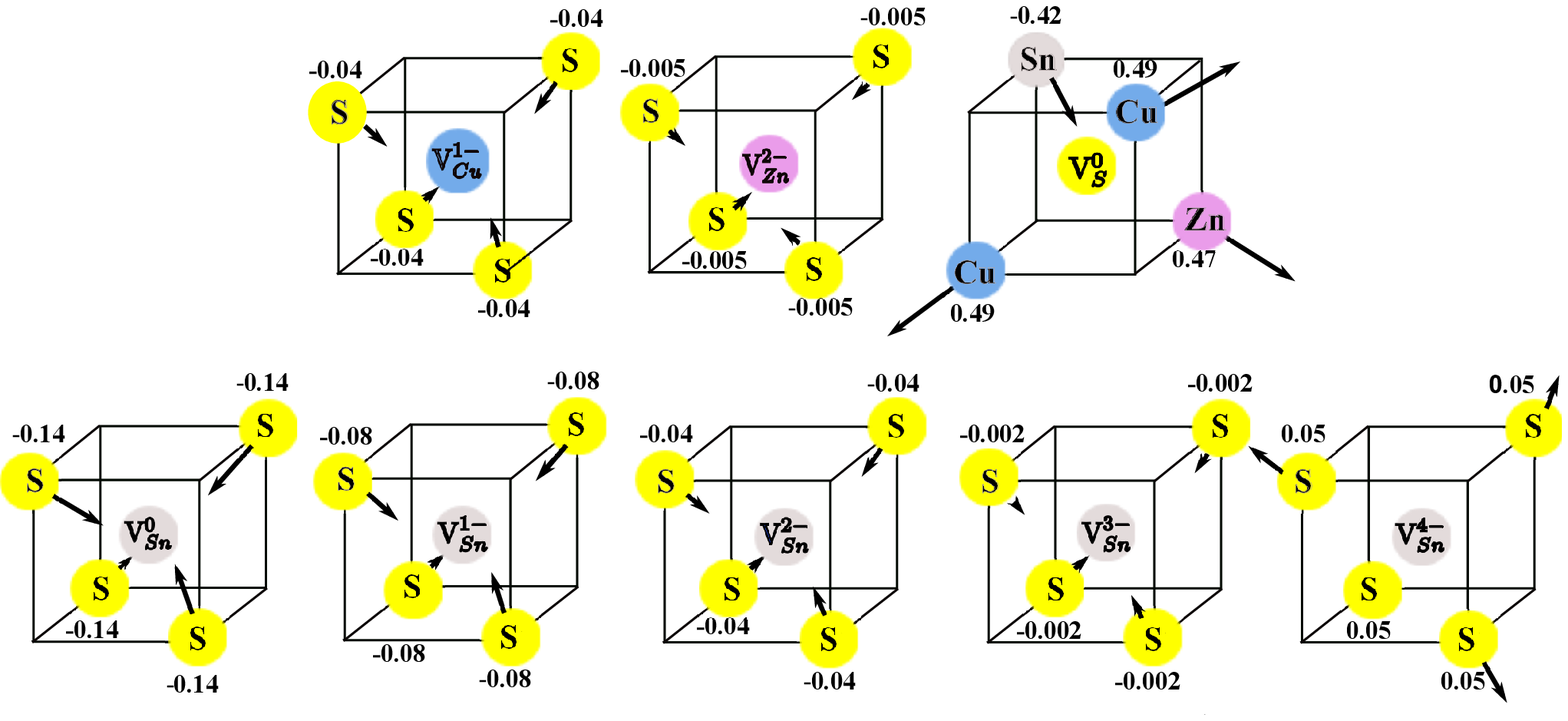}
\end{figure*}
\clearpage
\newpage


\begin{figure*}[htbp]
\center {$\Huge\textbf{Fig.2  \underline{Zhang}.eps}$}
\includegraphics[bb = 10 10 700 300, width=2\textwidth]{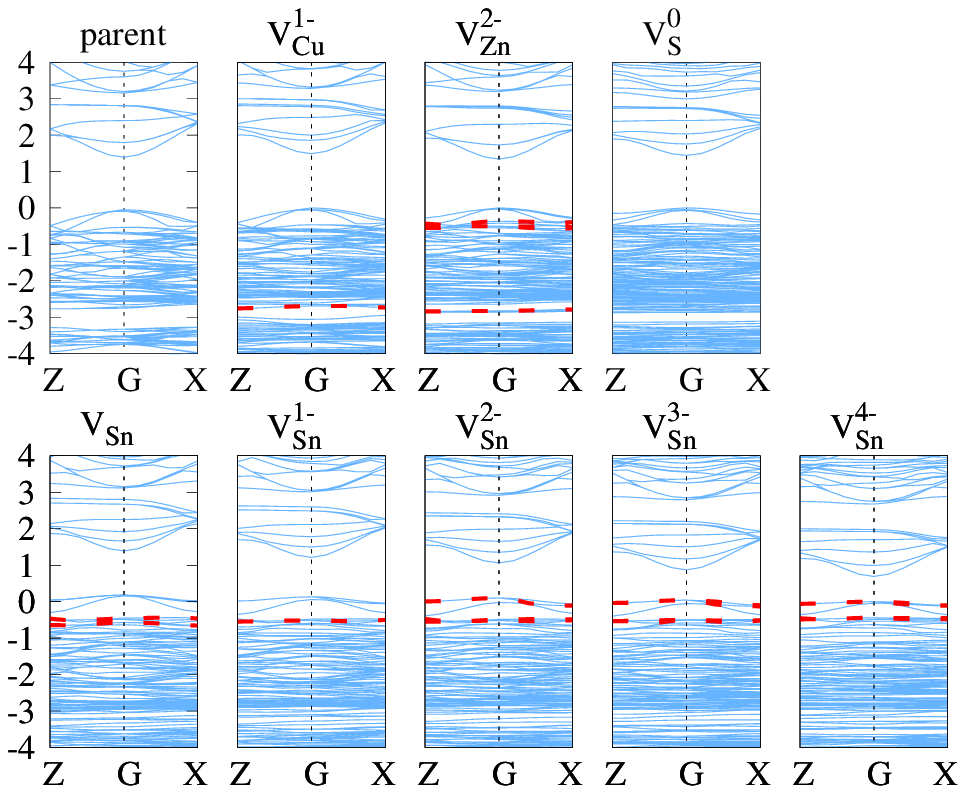}
\end{figure*}
\clearpage
\newpage


\end{document}